\documentclass{ws-procs9x6}

\begin{document}

\title{RECENT DEVELOPMENTS ON HADRON INTERACTION AND DYNAMICALLY GENERATED RESONANCES}

\author{E. Oset$^*$, M. Albaladejo and Ju-Jun Xie$^1$}

\address{Departamento de F\'{\i}sica Te\'orica and IFIC, Centro Mixto Universidad de Valencia-CSIC,
Institutos de Investigaci\'on de Paterna, Aptdo. 22085, 46071 Valencia,Spain\\
$^*$E-mail: oset@ific.uv.es\\
$^1$: Also Institute of Modern Physics, Chinese Academy of Sciences, Lanzhou 730000, China, and State Key Laboratory of Theoretical Physics, Institute of Theoretical Physics, Chinese
Academy of Sciences, Beijing 100190, China
}

\author{A. Ramos}

\address{Departament d'Estructura i Constituents de la Mat\`eria and Institut de
Ci\`{e}ncies del Cosmos, Universitat de Barcelona, Mart\'{\i} i Franqu\`es 1, E-08028
Barcelona, Spain\\
}

\begin{abstract}
In this talk I report on the
recent developments in the subject of dynamically generated resonances. In particular I discuss the $\gamma p \to K^0 \Sigma^+$ and  $\gamma n \to K^0 \Sigma^0$ reactions, with a peculiar behavior around the $K^{*0} \Lambda$ threshold,  due to a $1/2^-$ resonance around 2035 MeV. Similarly, I discuss a BES experiment, $J/\psi \to \eta K^{*0} \bar K^{*0}$ decay, which provides evidence for a new $h_1$ resonance around 1830~MeV that was predicted from the vector-vector interaction. A short discussion is then made about recent advances in the charm and beauty sectors.  
\end{abstract}


\bodymatter

\section{Introduction}\label{intro}
The unitary treatment of coupled channels with interaction kernels extracted from chiral Lagrangians has given rise to the chiral unitary approach that provides scattering amplitudes for hadron hadron interaction and, by looking at poles of the scattering amplitudes, bound states or resonances which we call dynamically generated \cite{review}.  The chiral Lagrangians are extrapolated to include vector mesons by means of the local hidden gauge Lagrangians \cite{hidden1} and then one can study vector-vector and vector-baryon interactions \cite{reportvector}. In the vector-vector sector this interaction has been shown to lead to many resonances \cite{gengvec}, some of which can be associated to known states, while a few others are predictions. One of these is a $h_1$  ($0^-(1^{+-})$) state which couples only to $K^* \bar K^*$. 

  In the vector-baryon sector, several resonances around 2000 MeV have been found in \cite{angelsvec} from the interaction of vectors with the members of the proton SU(3) octet. Some $1/2^-$ states appear from the interaction which can be associated to known states, but some are predictions. In fact, there are predictions of states around 2000 MeV that could be associated to states previously catalogued in the PDG, but which have disappeared in the latest edition of the Book. 

   In the charm and beauty sectors the proliferation of X,Y,Z mesonic states which do not fit the conventional charmonium spectrum have led to a plethora of works suggesting molecules, tetraquarks, and other exotic states \cite{slzhu}. The baryon sector has followed this trend with multiple suggestions of non conventional baryons \cite{zouhyper}. In this talk I will address some examples in these sectors. 

\section{Signature of an $h_1$ state in the  $J/\psi \to \eta K^{*0} \bar K^{*0}$ decay}

\begin{figure}[t!]\centering
\includegraphics[width=0.55\textwidth]{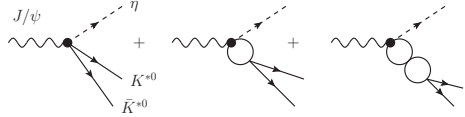}
\caption{Diagrammatic representation of the $J/\psi \to \eta
K^{*0}\bar{K}^{*0}$ decay. \label{feydgm}}
\end{figure}

\begin{figure}[t!]\centering
\includegraphics[width=0.55\textwidth]{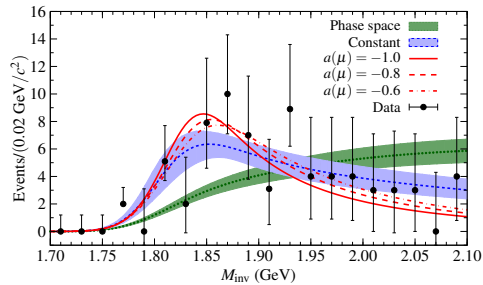}
\caption{(Color online) The $K^{*0}\bar{K}^{*0}$ invariant mass
spectrum of $J/\psi \to \eta K^{*0}\bar{K}^{*0}$ decay. The data points are taken from
Ref.~\cite{BESdata}. The different lines represent the output  for different approaches. The short-dashed line and the associated error band (light blue) represent the results of a constant potential. The (red) solid, long-dashed and dot-dashed lines represent the results for the potential of \cite{gengvec} with $a(\mu)=-1.0$, $-0.8$ and $-0.6$, respectively. Finally, the (green) dotted line, and the associated error band (dark green) is the prediction for phase space alone.\label{fig:dgdm}}
\end{figure}

We plot diagrammatically the process in Fig. \ref{feydgm}. If we produce an $\eta$ and a $K^* \bar K^*$ in the most favorable process involving L=0 in the vertex, then the quantum 
numbers of the $K^* \bar K^*$ pair are $I^G(J^{PC})=0^-(1^{+-})$. These are the quantum numbers of an $h_1$ resonance. The produced  $K^* \bar K^*$ will interact, which is depicted by the bubbles in the figure signifying the multiple interactions of the Bethe Salpeter equation. Then, if this interaction is strong enough to produce a resonance, as found in \cite{gengvec},  a peak will be produced in the invariant mass of a final $K^{*0} \bar K^{*0}$ state. This is what is seen in the experiment \cite{BESdata}, as shown in Fig. \ref{fig:dgdm}. In \cite{miguel} the evaluation is done for this invariant mass distribution and taking the input from \cite{gengvec}, up to a an arbitrary normalization, the curves of Fig. \ref{fig:dgdm} are obtained which provide a good reproduction of the data. A choice is made there of the subtraction constant of the loop function for the two 
$K^* \bar K^*$ mesons. With this subtraction constant demanded by experiment, one can go back to \cite{gengvec} and evaluate the $K^* \bar K^*$ scattering matrix with the $h_1$ quantum numbers, and one finds that the amplitude contains a pole corresponding to a resonance around 1830 MeV, in qualitative agreement with the prediction around 1800 MeV. The experimental data have served to obtain additional information that can make the prediction more precise. Hence the experiment provides evidence for a new $h_1$ state which is so far no catalogued.

\section{The $\gamma p \to K^0 \Sigma^+$ and  $\gamma n \to K^0 \Sigma^0$ }

\begin{figure}[htb]
\begin{center}
\includegraphics[width=0.38\textwidth]{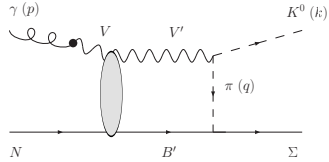}
\caption{Mechanism for the photoproduction reaction $\gamma N \to K^0\Sigma$.
The symbol $V$ stands
for the $\rho^0$, $\omega$ and $\phi$ mesons, while $V^\prime B^\prime$ denotes
the intermediate channel, which can be
$K^{*+}\Lambda$, $K^{*+}\Sigma^0$ or
$K^{*0}\Sigma^+$,
in the case of $\gamma p \to K^0\Sigma^+$, 
and $K^{*+}\Sigma^-$  or
$K^{*0}\Lambda$,
in the case of $\gamma n \to K^0\Sigma^0$, the $K^{*0}\Sigma^0$ 
intermediate channel does not
contribute due to the zero value of the
$\pi^0\Sigma^0\Sigma^0$ coupling at the Yukawa vertex. }
\label{fig:diag}
\end{center}
\end{figure}

In a recent experiment \cite{Ewald:2011gw}, the cross section for the $\gamma p \to K^0 \Sigma^+$ reaction exhibits a curious behavior close to the $K^* \Lambda $ threshold. The cross section suddenly drops and the differential cross sections become isotropic. The cross section is not reproduced by any of the standard models MAID, SAID, etc. A hint was given in \cite{Ewald:2011gw} that this could be due to the role played by intermediate states of vector-baryon. A materialization of this idea was recently made in 
\cite{angelsgam}. In this work the mechanism for $K^0 \Sigma^+$ photoproduction was taken as depicted in Fig. \ref{fig:diag}. The photon gets converted into a vector meson according to the vector dominance hypothesis \cite{hidden1}. Then the vector and the nucleon interact, as done in \cite{angelsvec} and finally, a vector and a baryon produce the final $K^0 \Sigma^+$ state via the exchange of a pion. In the energy region of interest the vector-baryon states of relevance are  $K^{*+}\Lambda$, $K^{*+}\Sigma^0$ or $K^{*0}\Sigma^+$ and we take all of them into consideration. We then repeat the same with the    $\gamma n \to K^0 \Sigma^0$ reaction.

\begin{figure}[h]
\begin{center}
\includegraphics[width=0.55\textwidth]{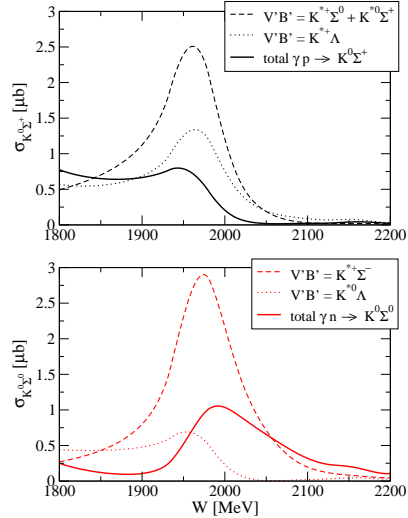}
\caption{Contributions to the $\gamma p \to K^0 \Sigma^+$ (upper panel) and 
$\gamma n \to K^0 \Sigma^0$ (lower panel) cross sections. }
\label{fig:gn_gp}
\end{center}
\end{figure}

    The cross section for the reaction shows up a peak due to a $N^*$ dynamically generated resonance around 2000 MeV \cite{angelsvec}, but the striking thing is that there is a destructive interference between the intermediate $K^* \Sigma$ and $K^* \Lambda$ channels such that the cross section for the proton falls down precisely where the individual contributions have the peak, as can be seen in Fig. \ref{fig:gn_gp}.  On the other hand, in the case of a neutron target, the $K^* \Lambda$ contribution is weaker and the peak of the cross section is still seen after the interference, such that the cross section on the neutron develops a peak where the cross section on the proton falls down. This is a net prediction of the theory which would be very interesting to be observed. Although with some difference, since we have more possibilities in the intermediate states, these findings are a reminiscence of those for $\eta$ photoproduction on protons and neutrons found in \cite{misha}, where a peak seen in the photoproduction on the neutron targets was interpreted as a consequence of the different interference of the $K \Sigma$ and $K \Lambda$ channels. In the case of $\gamma n$, the intermediate state for 
$K \Lambda$  is $K^0 \Lambda$ and the photon does not couple to the neutral $K^0$, hence there is only contribution from the $K \Sigma$ channel in that case.

\begin{figure}
\begin{center}
\includegraphics[width=0.50\textwidth]{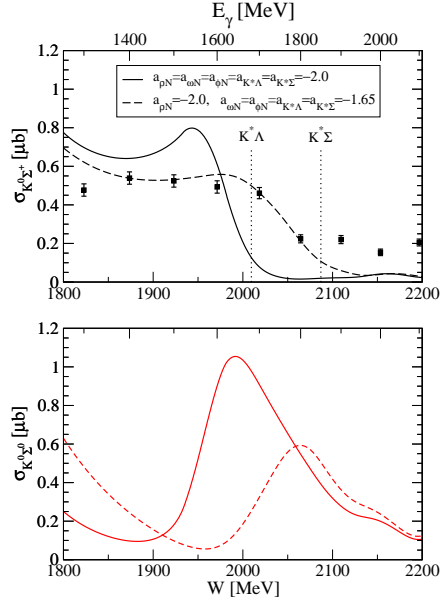}
\caption{Upper panel: Comparison of the $\gamma p \to K^0 \Sigma^+$ cross
section, obtained with two parameter sets, with the CBELSA/TAPS data of
Ref.~\cite{Ewald:2011gw}. The downfall of the cross section allows one to redefine
the parameters of the model and give a better prediction for the position of the
resonance.
Lower panel: Predictions for the $\gamma n \to K^0 \Sigma^0$ cross section using
two parameter sets. }
\label{fig:sigma}
\end{center}
\end{figure}

With the input used in \cite{angelsvec} we find the results of Fig. \ref{fig:sigma} shown by the solid line. The fall down appears but is displaced with respect to the experimental data. This serves us to fine tune the subtraction constant in the meson baryon loop function. Once this is changed, the agreement with data is then good. With the new parameters we then evaluate the cross section for the neutron target and we find a peak in the cross section at 2075 MeV, which is a prediction. Furthermore, with these parameters we can now evaluate the poles of the vector-baryon amplitude, finding one around 2030~MeV. Thus, the data can be interpreted as the effect of a dynamically generated resonance ($1/2^-, 3/2^-$) at 2030 MeV. 

\section{Dynamically generated states in the charm and beauty sectors}

This topic has generated much activity \cite{Kolomeitsev:2003ac,Hofmann:2003je,Guo:2006fu,Gamermann:2006nm,Gamermann:2007fi,
Hofmann:2005sw,Mizutani:2006vq,raquelxyz,Garcia-Recio:2013gaa,GarciaRecio:2012db,Romanets:2012hm} and recent short reviews can be seen in \cite{slzhu} and \cite{zouhyper}.  We shall address here some recent work on the topic.

    In \cite{wuprl,wu} baryon states of hidden charm were investigated using an extrapolation to SU(4) of the results of the local hidden gauge in SU(3). Actually the dominant terms in the potential come from the exchange of light vectors where the heavy quarks play the role of spectators. In this case the results can be obtained from a mapping of SU(3). It has been found recently \cite{xiaojuan,xiaoyo,xiaoaltug} that these results are consistent with heavy quark spin symmetry \cite{isgurwise,Neubert:1993mb}. 

In \cite{wu,wuprl} one obtains a $N^*$ and a $\Lambda ^*$ states in the pseudoscalar-baryon and vector-baryon cases. The nucleon state, with mass around 4260 MeV couples mostly to  $\bar{D} \Sigma_{c}$, while there are two $\Lambda$, one around 4200 MeV, which couples mostly to 
$\bar{D}_{s} \Lambda^{+}_{c}$  and $\bar{D} \Xi_{c}$, and another one at 4400 MeV, which couples mostly to  $\bar{D} \Xi'_{c}$. Similar states are found substituting the $\bar D$ by $\bar D^*$ displaced by about 150 MeV with respect to the PB states.  These states are also found to have relatively narrow widths of about 50 MeV. It is thus quite surprising to find $N$ and $\Lambda$ states with such a large mass and a small width. The reason is that they can decay to lighter channels of PB or VB, but the transitions are penalized by forcing the exchange of a heavy vector that reduces the strength of the matrix element. 

  In a more recent paper, these ideas are generalized to the beauty sector \cite{Wu:2010rv}
and again the states obtained have masses of around 11000 MeV.

In \cite{xiaojuan} baryon states of hidden charm are studied from the perspective of heavy quark spin symmetry using dynamics of the local hidden gauge. The results of \cite{wu} are reproduced and extended to other combinations of pseudoscalar-baryon or vector-baryon. In \cite{xiaoyo} this is extended to the beauty sector, and one again, the results of   \cite{Wu:2010rv} are reproduced and more states are predicted from different combinations of meson-baryon channels. An extrapolation of these ideas is also done to the meson-meson  sector, where meson states of hidden beauty are generated \cite{xiaoaltug}. A fair amount of states is predicted which we hope will be found experimentally in the future.

   Another source of predictions in this sector comes from the use of Heavy quark spin symmetry, using some phenomenological input, where results qualitatively similar to those reported before are also obtained  \cite{HidalgoDuque:2012pq,Nieves:2012tt,Guo:2013sya}. A variety of methods are also used in other works 
\cite{Liu:2010xh,Zhang:2006ix,zzy1,Cleven:2013mka}. The list of work done in this issue is long giving evidence of a thriving field.

\section{Conclusions} 

The chiral unitary approach has proved very efficient to study the interaction of hadrons, and sometimes the scattering matrices show poles that are consequence of the unitarization with the potentials obtained from the chiral Lagrangians, signalling the existence of
dynamically generated states, which are a kind of molecular states of some hadrons. Mounting experimental information is giving support to the nature of these states by finding some of the states predicted that had not been reported before. On the other hand, and curiously, it has been in the charm and beauty sectors that many states have been found that do not fit in the ordinary scheme of the quark model and call for more complex structures, some of them clearly of molecular nature. The theoretical effort in these sectors is impressive, as well as the amount of states predicted. With the continuing observation of new states in these sectors in the different facilities around the world, we can only hope that these predicted states are gradually found and we deepen our understanding of hadron physics.

\section*{Acknowledgments}

This work is partly supported by the Spanish Ministerio de Econom\'\i a y Competitividad and
European FEDER funds under the contract number FIS2011-28853-C02-01 and  FIS2011-28853-C02-02, and the Generalitat
Valenciana in the program Prometeo, 2009/090. This work is also partly supported by the National Natural Science Foundation of China under grant 11105126. We acknowledge the support of the European
Community-Research Infrastructure Integrating Activity Study of Strongly Interacting Matter
(HadronPhysics3, Grant Agreement n. 283286) under the Seventh Framework Programme of the EU. 

\bibliographystyle{ws-procs9x6}
\bibliography{ws-pro-sample}

\end{document}